\documentclass[10pt,a4paper,twoside]{article}
\usepackage{epsfig}
\usepackage{baltlat6}
\usepackage{array}
\usepackage{here}
\begin{document}
\ \
\vspace{0.5mm}
\setcounter{page}{1}

\titlehead{Baltic Astronomy, vol.\,23, 260-266, 2014}

\titleb{MEDIUM RESOLUTION SPECTROSCOPY AND CHEMICAL\\
  COMPOSITION OF GALACTIC GLOBULAR CLUSTERS}

\begin{authorl}
\authorb{D.A. Khamidullina}{1}, 
\authorb{M.E. Sharina}{2},
\authorb{V.V. Shimansky}{1}, 
\authorb{E. Davoust}{3}
\end{authorl}

\begin{addressl}
\addressb{1}{Kazan Federal University, Kremlevskaya 18, Kazan, 420008, Russia;\\ hamidullina.dilyara@gmail.com}
\addressb{2}{Special Astrophysical Observatory, Russian Academy  of Sciences,\\
N. Arkhyz, KChR, 369167, Russia; sme@sao.ru}
\addressb{3}{IRAP, Universit\'e de Toulouse, CNRS, 14 avenue E. Belin, F-31400, France;\\ edavoust@irap.omp.eu}
\end{addressl}

\submitb{Received: 2014 September 15; accepted: }

\begin{summary} We used integrated-light medium-resolution spectra of six Galactic 
globular clusters and model stellar atmospheres to carry out population
synthesis and to derive the chemical composition and age of the clusters. We
used medium-resolution spectra of globular clusters published by Schiavon et
al. as well as our long-slit observations with the 1.93 m telescope of the Haute
Provence Observatory. The observed spectra were fitted to the theoretical ones
interactively. As an initial approach, we used masses, radii and
 log\,$g$ of stars in
the clusters corresponding to the best fitting isochrones in the observed 
color-magnitude diagrams. The computed synthetic blanketed spectra of stars were
summed according to the Chabrier mass function. To improve the determination
of age and helium content, the shape and depth of the Balmer absorption
lines was analysed. The abundances of Mg, Ca, C and several other elements
were derived. We achieved a reasonable agreement with the literature data
both in chemical composition and in age of the Galactic globular clusters. 
Our method might be useful for the development of stellar population models 
and for a better understanding of extragalactic star clusters.

\end{summary}


\resthead{Chemical composition of globular clusters}
{Khamidullina et al.}

\sectionb{1}{INTRODUCTION} 
Star clusters in galaxies of different morphological types have long been considered as the
 best representatives of so-called simple stellar populations, i.e. those consisting of stars 
with the same age and metallicity. Recent deep photometric and high-resolution spectroscopic 
studies of Galactic globular clusters (GGCs) have shown that these old stellar systems are not
 as simple as previously thought. Globular clusters (GCs) may consist of stars of different
 chemical composition and show element abundance anti-correlations (Gratton et al. 2012).
 Understanding the real accuracy of the elemental abundance and age determination based on 
medium-resolution spectra is important for studies of extragalactic GCs. 

Our idea was to use models of stellar atmospheres for the population synthesis
and determination of element abundances, because this method is not limited in
the abundance patterns and atmospheric parameters, as opposed to 
methods using empirical stellar libraries.
We understand that our method also has disadvantages. Computations of theoretical spectra use
 incomplete lists of atomic and molecular lines and inaccurate atomic constants (e.g. Coelho
 2014). This is why a comparison of our derived parameters with the literature
 is very important for understanding the quality of the spectral fitting.

Catalogues by Harris (1996), Borkova \& Marsakov (2005) contain a summary of 
 fundamental parameters of our sample GGCs: NGC~104, NGC~6121, NGC~6205, NGC~6218, NGC~6838,
 NGC~7078. We selected objects with different metallicities and horizontal branch (HB)
 morphologies to check how these characteristics influence the derived parameters.

Technical details of the 1.93m telescope observations and the methodological subtleties of
 our analysis are discussed in our previous papers (Sharina et al. 2013, 
Khamidullina et al. 2014). Here we briefly describe the main principles of our analysis 
and present the results using NGC7078 as an example. 

\sectionb{2}{DATA REDUCTION AND ANALYSIS}

We took spectra of NGC104, NGC6121 and NGC6218 from the library of Schiavon (2005). 
The long-slit spectra of the other three clusters (NGC6205, NGC6838 and NGC~7078) 
were obtained during our observations with the CARELEC spectrograph of the 1.93-m telescope 
in the Haute Provence Observatory in July 2010. We used a grating with 300 grooves/mm and 
the slit width 2". The seeing during our observations was $2.5 - 3\arcsec$.
The spectral range was 3700 - 6800 \AA~.
Each GC was observed using several fixed slit orientations centred at the core 
of the cluster with exposures of $1200 \div 1800$ sec. Thus we were able to remove 
background and foreground objects from the two-dimensional spectra. 
This would not be possible in the case of scanning the clusters with a moving, or rotating slit 
(Schiavon et al., 2005).
A detailed description of the observational process and instrumental 
conditions was published by Khamidullina et al. (2014). The mean
spectral resolution in the middle of the spectral range was FWHM$\sim 5$\AA. 
The primary reduction of the 1.93-m telescope data was done using the 
{\sc MIDAS} Banse et al. (1983) and {\sc IRAF}{\footnote {http://iraf.noao.edu/}}
software packages. 

We consider spectroscopic and photometric observational data and theoretical model capable to explain them. 
The method is based on the comparison of the observed spectra to the theoretical ones computed using 
synthetic spectra of stars of different masses, $log~g$ and $T_{eff}$.
 The stellar parameters were set by the best isochrone fitting of the cluster CMD.
We used the isochrones of Bertelli et al., 2009~{\footnote {http://stev.oapd.inaf.it/YZVAR/)}}, 
because these models contain the Horizontal branch (HB) and Asymptotic branch (AGB) stages 
of evolution and a wide range of model parameters (metallicity, $\alpha$-element ratio, age and helium content).
We used deep stellar photometry data from Piotto et al. (2002) and Sarajedini et al. (2007) to be compared with 
the isochrones. A number of stars of a particular mass bin was computed according to
the Chabrier mass function (Chabrier, 2005).

Ages and abundances of different chemical elements were derived via 
fitting of the observed spectra to computed theoretical ones.
The synthetic spectra were computed using the {\sc SPECTR} software package (Shimansky et al. 2003) 
by interpolating the model grid of Castelli \& Kurucz (2003) 
(see Shimansky et al. 2003 and Sharina et al. 2014 for a full description of the programs and atomic constants).
Plane-parallel, hydrostatic model stellar atmospheres were calculated
for a given  set of parameters (T$_{eff}$, $log~g$, $[M/H]$) with and without atomic and molecular lines. 
The solar chemical abundances were taken from Asplund et al. (2006) 
for Fe, C, N, and O and from Anders \& Grevesse (1989) for all the other elements.
The profiles of the Balmer lines were calculated according to the  
theories of Vidal, Cooper \& Smith (1973) and Griem (1960).
We used the line lists of Kurucz (1994) and Castelli \& Kurucz (2003), and 28
bands of 10 molecules (VO, TiO, SO, SiO, NO, MgO, MgH, CO, CN, AlO) 
computed with the theory of Nersisyan et al. (1989) 
and kindly provided by Ya.V. Pavlenko. Abundances derived using strong dominant lines 
of Ca, Mg, Fe, CH, Na, Al, Ba, Sr and all lines with $\lambda > 5300$ \AA~ are differential, 
because we used empirical oscillator strengths $gf$ from Shimanskaya et al. (2011). 
Theoretical $gf$ were used for the other lines. This may lead to under-estimation 
of the abundances by $\sim0.07$ dex.

We emphasize that the analysis of medium-resolution spectra permits to use only wide blends 
of lines and molecular bands with widths $\Delta \lambda \ge 5$\AA, but not individual lines of chemical elements. 
Thus, to reach reliable results one has to consider high signal-to-noise spectra ($S/N \ge 100$) containing several 
absorption line features of the element in a wide spectral range.

Setting the continuum level represents the major difficulty in the process of fitting synthetic spectra to 
low resolution spectra. We fitted the pseudo-continuum to the whole spectrum using the {\sc ULySS} 
program\footnote{http://ulyss.univ-lyon1.fr} (Koleva et al. 2009). 
Multiplicative and additive polynomials were applied to 
the observed spectrum to bring it in agreement with the model one.
After the procedure, there were still uncorrected high-frequency variations of the
pseudo-continuum. However, they did not introduce any systematics in our results,
because we estimated the abundances of chemical elements having many intense (dominant) spectral lines in the 
studied wavelength range. The atmospheric parameters and abundances were adjusted using pixel-to-pixel spectral 
fits of many metal absorption lines over a wide spectral range. 

The contribution of different types of stars to the integrated light of an old GC is considered by 
Khamidullina et al. (2014). Hot blue HB stars contribute up to 40$\%$ spectral intensity in the blue range. 
On the other hand, red giants dominate in the red part. They produce most of molecular bands and lines of metals. 

\sectionb{3}{RESULTS}
We test our method using old Galactic clusters. Here we demonstrate just one example:
one of the most metal-poor GCCs NGC~7078 (M15)([Fe/H]$=-2.26$ dex (Harris (1996), Preston et al.(2006) 
and references therein). This GC is one of the most compact ones in the Milky Way. 
It belongs to the young Galactic halo according to Borkova \& Marsakov (2000). 
NGC~7078 was observed by us and by Schiavon et al. (2005).
We thus have a possibility to compare the spectral fitting results obtained using one method, 
but different spectra with a spectral resolution of $FWHM = 3$ and 5\AA~.

This GC is of intermediate HB morphology. An extended blue tail of the HB and a number of 
blue straggler stars are seen on its CMD. 
The HB of the cluster contains a large number of stars in its blue and middle parts.
Examples of isochrone fitting for NGC~7078 are shown in Fig.~1. We used the stellar photometric 
results by Piotto et al. (2002).
The isochrones of different age with all other parameters close to the 
ones derived by Vandenberg et al. (2013) for NGC~7078 are shown in the left and right panels. 
\begin{figure}[!tH]
\vbox{
\vspace{-3mm}
\centerline{\psfig{figure=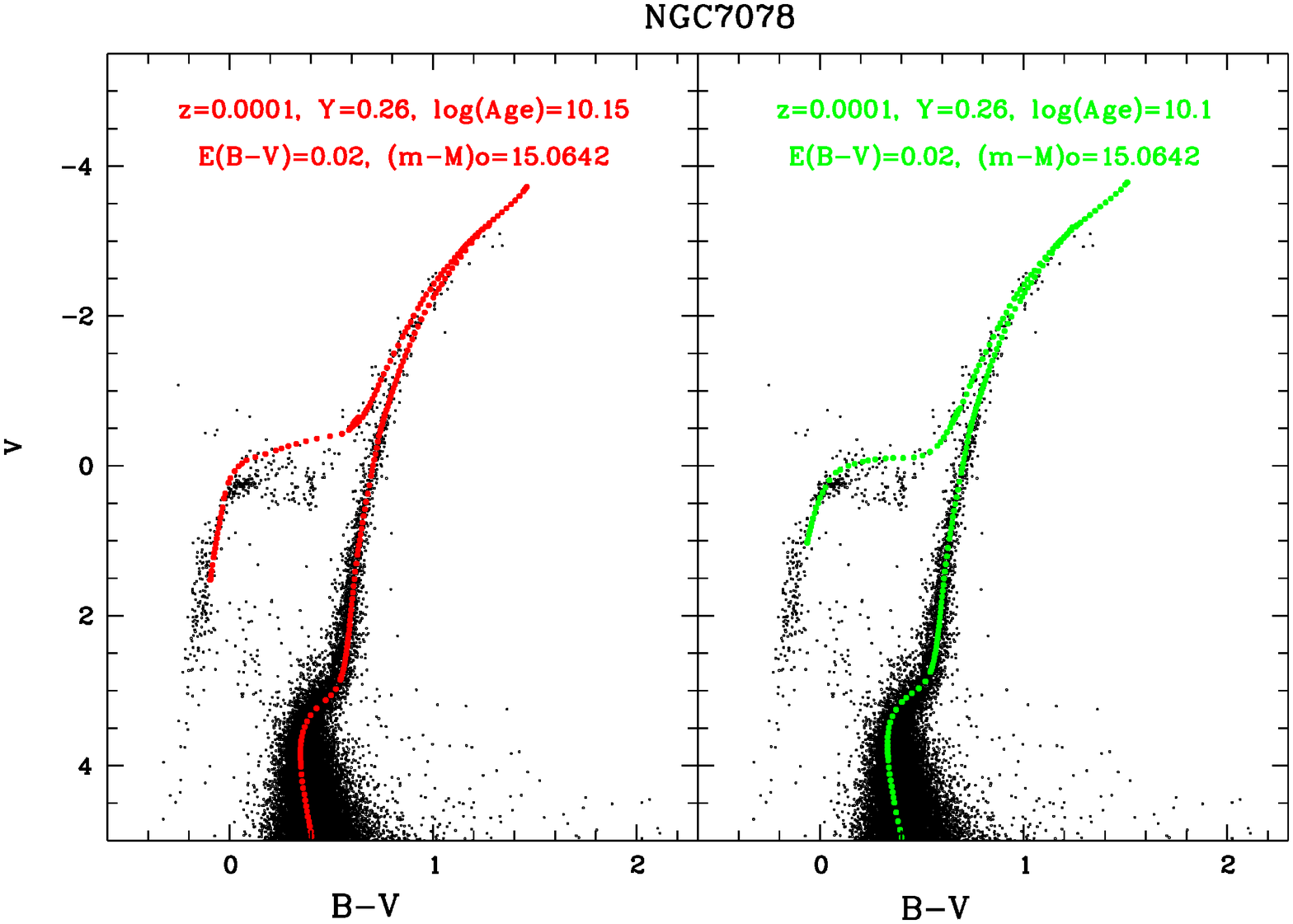,width=100mm,angle=0,clip=}}
\vspace{1mm}
\captionb{1}
{$I$ vs. $V-I$ diagram for NGC~7078 (Piotto et al., 2002) with isochrones
from Bertelli et al. (2009) overplotted. Left: isochrone with $Log Age=10.15$, $Y=0.26$, $Z=0.001$, 
values used by us for the spectral fitting. Right: isochrone with $Log Age=10.10$, $Y=0.26$ and $Z=0.0001$, 
values used by Vandenberg et al. (2013).}}
\end{figure}

The comparison of the observed spectrum of NGC~7078 to the computed synthetic one 
is shown in Fig.~2.
It appears that the choice of the older age gives better results in the sense that the depth and 
shape of the Balmer absorption lines in the synthetic spectrum fit the observed lines better if we choose 
$log~age = 10.15$. 
\begin{figure}[!tH]
\vbox{
\vspace{-10mm}
\centerline{\psfig{figure=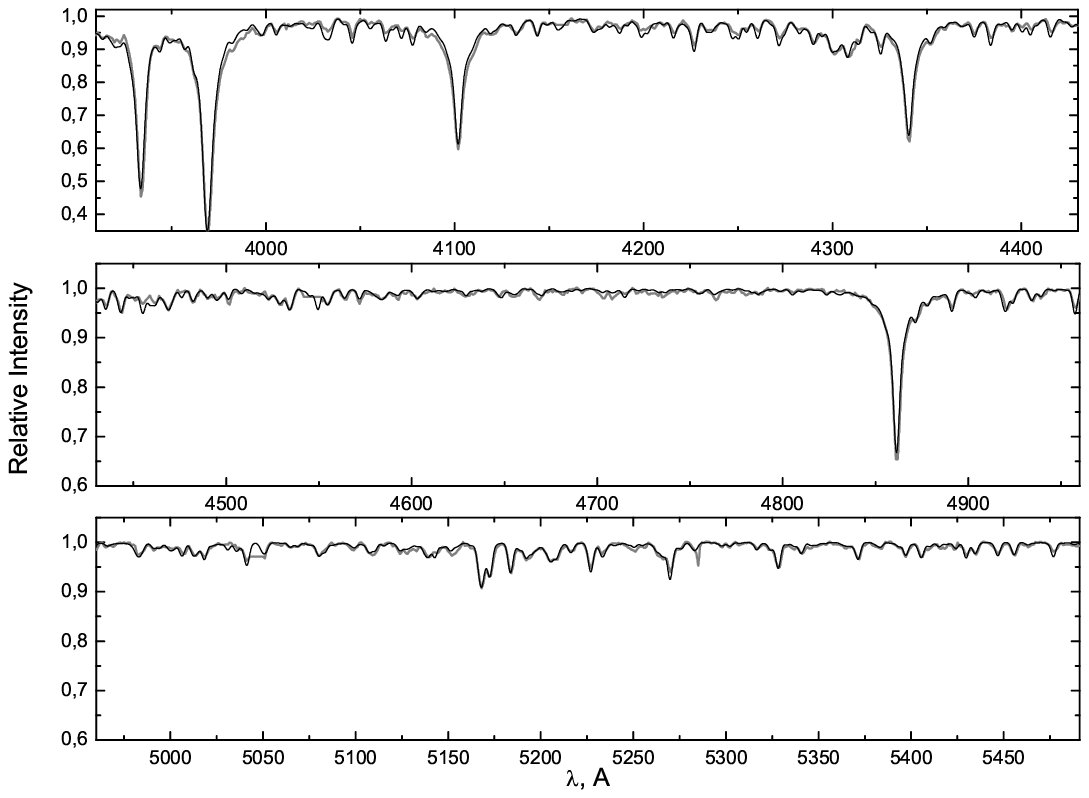,width=140mm,angle=0,clip=}}
\vspace{1mm}
\captionb{2}
{Computed theoretical spectrum (black line) compared with the observed one of NGC~7078 (grey line).}}
\end{figure}

The element abundances for NGC~7078 and other eight GGCs from this 
paper and from Khamidullina et al. (2014) are shown in Fig.~3, where they are compared 
to the literature abundances summarised by Pritzl et al. (2005) (open circles) and Roediger et al. (2014) (dots). 
The abundances of the elements with dominant lines and molecular bands were estimated with accuracies 
$\Delta [X/H] \sim 0.1 \div 0.2$ dex. These are: $Fe$, $Ca$, $Mg$. The line profiles of $Ti$, $Cr$ and $Mn$ 
are weak and blended. The accuracies of their abundances are $\sim 0.2 \div 0.3$ dex. Some elements have 
no spectroscopic features distinguishable at our resolution. However, they influence the ionisation 
equilibrium of other elements. These are, for example, $O$, $Al$, $Si$, $V$, $Ni$. Uncertainties of 
the derived abundances of these elements are larger than $0.3$ dex. The resonance lines of $Na$ are 
strongly distorted by the interstellar extinction and should be included in the last group.
It is worth noting that considerable differences between our abundances and the literature ones may 
arise for some elements with weak absorption lines due to strong statistical noise in some parts of 
our spectra influenced by sky emission and absorptions lines, or by cosmic particles. It is surprising 
that, in the case of NGC~7078, the iron and magnesium abundances differ strongly from the literature ones. 
This effect exists whether we use our spectrum or the data from the library of Schiavon et al.(2005). 
It is possible also that a bright star with a velocity close to that of NGC~7078, but a different 
chemical composition is projected on the centre of the cluster. 

Considerable systematic differences between our values and the literature ones were obtained for the 
abundances of $C$ and $N$ (Fig.~3). A possible explanation of this fact is the following. We analyse 
the chemical composition of the whole GC, and not only of the most luminous red giant branch stars, 
as is done in high resolution spectroscopic studies.
The outer atmospheric layers of such stars contain products of the CNO cycle.
This effect leads to deriving an enhanced nitrogen abundance and a reduced carbon one.


\begin{figure}[!tH]
\vbox{
\centerline{\psfig{figure=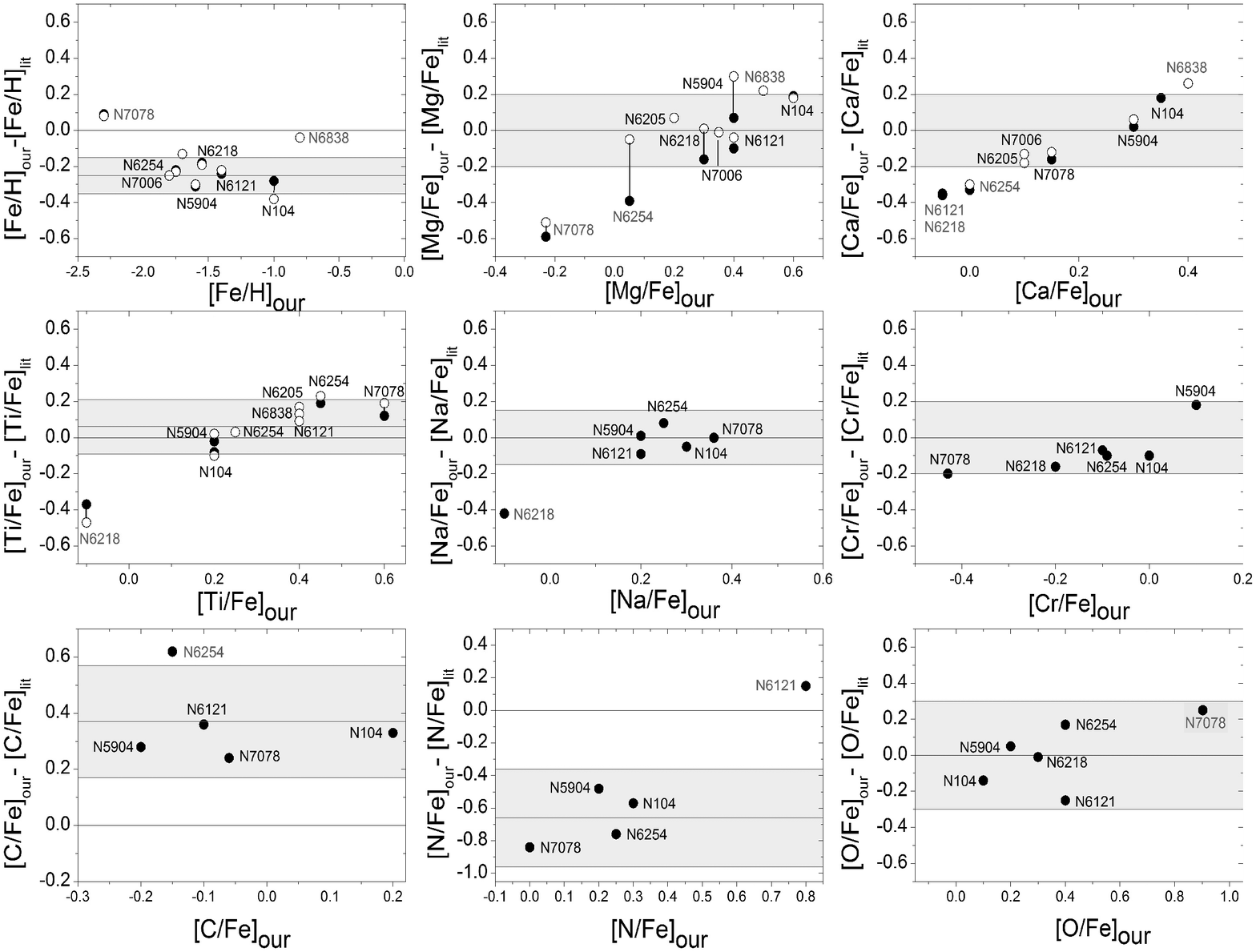,width=135mm,angle=0}}
\vspace{1mm}
\captionb{3}
{Comparison of our derived elemental abundances with the values from the papers of Pritzl et al. (2005) 
(open circles) and Roediger et al. (2014) (dots).}}
\label{compar}
\end{figure}



\sectionb{3}{CONCLUSION}
We used models of stellar atmospheres to synthesize stellar populations in six GGCs and compared our 
results to high-resolution spectroscopic and deep photometric data from the literature.
A good agreement was reached for the old clusters observed spectroscopically at a resolution $FWHM \ge 5$\AA~, 
in a wide spectral range $\Delta \lambda \ge 2000$ \AA~ and having signal-to-noise $S/N>100$.

\thanks{We acknowledge attribution of the Russian Federal Basic Research regional grant number 
14-02-96501-r-ug-a which however was not received because of the Karachai-Chekessian republic. 
SVV acknowledges a grant RFBR13-02-00351. We acknowledge the usage of the SIMBAD database operated 
at the CDS (Strasbourg, France), and Google. We thank organisers of the conference "Modern stellar astronomy", 
where this talk was presented in 2014.}

\References

\refb Asplund M., Grevesse N., Sauval A. J., Scott P., 2009, ARA\&A, 47, 481

\refb Anders E., Grevesse N., 1989, Geochim. Cosmochim. Acta, 53, 197

\refb Banse K., Crane Ph., Ounnas Ch., and Ponz D., 1983, in: Proceedings of the DECUS Europe Symposium, Zurich, Switzerland, Digital Equipment Computer Users Society, p. 87 

\refb Bertelli G, Nasi E, Girardi L. and Marigo P., 2009, A\&A, 508, 335

\refb Borkova T. V., Marsakov V. A., 2005, Astronomy Reports, 49, 405

\refb Borkova T. V., Marsakov V. A., 2000, Astronomy Reports, 44, 665

\refb Castelli F., Kurucz R. L., 2003, Proceedings of IAU Symp. 210: 
Modeling of Stellar Atmospheres, Eds. N. Piskunov et al., Poster A20

\refb Chabrier G., 2005, in The Initial Mass Function 50 years later. Eds. E. Corbelli et al. (Springer, Dordrecht), 327, 41

\refb Coelho, P. R. T., 2014, MNRAS, 440, 1027

\refb Gratton R., Sneden C., Carretta E., 2004, ARA\&A 42, 385

\refb Grevesse N., Sauval A.J., 1998, Space Science Reviews 85, 161

\refb Griem H.R., 1960, ApJ, 132, 883

\refb Harris W.E., 1996, AJ, 112, 1487 (http://www.physics,mcmaster.ca/Globular.html)

\refb Khamidullina D.A., Sharina M.E., Shimansky V. V., Davoust E., 2014, Astrophysical Bulletin, accepted

\refb Koleva, M.; Prugniel, Ph.; Bouchard, A.; Wu, Y., 2009, A\&A, 501, 1269

\refb Kurucz R.L. 1995, SAO CD-ROMs No. 23, Cambridge, MA, USA

\refb Kurucz R.L. 1994, SAO CD-ROMs No. 19-22, Cambridge, MA, USA

\refb Kurucz R.L. 1993, SAO CD-ROMs No. 1-18, Cambridge, MA, USA

\refb Nersisyan S. E., Shavrina A. V., Yaremchuk A. A., 1989, Astrofizika, 30, 247

\refb Piotto G, King I.R, Djorgovski S.G et al., 2002, A\&A, 391, 945

\refb Preston G.W., Sneden C., Thompson I.B., Shectman S.A., Burley G.S., 2006, AJ, 132, 85

\refb Pritzl B.J. and Venn K.A, 2005, AJ, 130, 2140

\refb Roediger J.C., Courteau S., Graves G. and Schiavon R.P, 2014, ApJS, 210, 10

\refb Sarajedini A., Bedin L.R., and Chaboyer B. et al., 2007, AJ, 133, 1658

\refb Schiavon R.P., Rose J.A., Courteau S. and MacArthur L.A., 2005, ApJS, 160, 163

\refb Sharina M.E., Donzelli C., Shimansky V. V., Davoust E., Charbonnel C., 2014, A\&A 570, 48

\refb Sharina M.E., Shimansky V. V., Davoust E., 2013, Astronomy Reports, 57, 410

\refb Shimansky V.V., Borisov N. V., Shimanskaya N. N., 2003, Astronomy Reports 47, 763

\refb Shimanskaya N. N., Bikmaev I. F., Shimansky V. V., Astrophysical Bulletin 66, 332

\refb VandenBerg D.A., Brogaard K., Leaman R. and Casagrande L., 2013, AJ, 775, 134

\refb Vidal C. R., Cooper J., Smith E. W., 1973, A\&A Suppl. Ser. 25, 37

\end{document}